\documentclass{aastex631}
\usepackage{CJK}

\usepackage{newtxtext,newtxmath}
\usepackage{physics}
\usepackage{graphicx}	% Including figure files
\usepackage{amsmath}	% Advanced maths commands
\usepackage{multirow}
\usepackage{hyperref,url}
\usepackage{xcolor}
\usepackage{soul}  % highlight the modifications
\newcommand{\kmspMyr}{\,km\,s$^{-1}$\,Myr$^{-1}$}

\begin{document}
\begin{CJK*}{UTF8}{gbsn}
\title{Impeding Turbulence Decay in Self-gravitating Cloud Cores}

\author[0009-0007-0109-743X]{Shibo Yuan (袁识博)}
\affiliation{Department of Physics, The Chinese University of Hong Kong, Shatin, Hong Kong}

\author[0000-0003-2641-9240]{Hua-Bai Li ({\CJKfamily{bsmi}李華白)}}
\affiliation{Department of Physics, The Chinese University of Hong Kong, Shatin, Hong Kong}

\correspondingauthor{Hua-Bai Li}
\email{hbli@cuhk.edu.hk}

\begin{abstract}

Turbulence governs the fragmentation of molecular clouds and plays a pivotal role in star formation. The persistence of observed cloud turbulence suggests it does not decay significantly within the turnover timescale, implying a recurrent driving mechanism. Although ubiquitous self-gravity is a plausible driver, MHD simulations by \citet{ostriker2001density} demonstrated that self-gravity alone does not modify the global turbulence decay rate. In this study, we demonstrate that the dominant diffuse volume of a cloud dictates its overall decay rate, while individual dense cores can maintain near-zero decay rates. Crucially, this phenomenon is absent in control simulations excluding self-gravity. This discrepancy cannot be attributed to contamination of turbulent velocities by core contraction, as most cores in our simulations remain in a quasi-equilibrium state. Our analysis reveals that the gravitational potential energy released during core formation\textemdash not necessarily driven by self-gravity but also by turbulent compression\textemdash is sufficient to sustain the observed turbulence levels within cores.
\end{abstract}
	

%%%%%%%%%%%%%%%%%%%%%%%%%%%%%%%%%%%%%%%%%%%%%%%%%%

%%%%%%%%%%%%%%%%% BODY OF PAPER %%%%%%%%%%%%%%%%%%

\section{Introduction}

Magnetic field and turbulence are crucial in the star formation process within molecular clouds, as indicated by various studies \citep{larson1981turbulence, shu1987star, mckee1993magnetic, crutcher2012magnetic, li2021magnetic,pattle2022magnetic}. Magnetic fields can guide gas flows into anisotropic turbulence, while turbulence, in turn, can amplify or distort the magnetic fields. This complex interplay results in the formation of magneto-hydrodynamic (MHD) turbulence \citep{li2017taich5,li2021magnetic}. Extensive research has been conducted on the MHD turbulence within self-gravitating clouds, examining parameters such as the dissipation timescale \citep{ostriker2001density}, degree of velocity anisotropy \citep{Otto_2017}, turbulent energy solenoidal ratio, and injection scale of gravity-driven turbulence \citep{Federrath2011ApJ}, as well as the saturation level of turbulence during gravitational collapse \citep{Higashi2022ApJ}.
This work specifically investigates the decay of turbulence, reconciling previous findings and exploring density-dependent effects.

Without non-thermal support, molecular clouds should collapse and form stars in a small fraction of their observed lifetime. Meanwhile, supersonic turbulence is commonly observed in molecular clouds \citep{larson1981turbulence, ossenkopf2002turbulent, heyer2004universality}, prompting the proposal of hydrodynamic (HD) turbulence as a support mechanism. However, HD turbulence was expected to decay within the cloud's free-fall timescale if not continuously driven. Researchers have used MHD simulations to explore whether magnetic fields could slow down turbulence decay, by studying the power-law relation $E_\text{turb}\propto t^\eta$, where $E_\text{turb}$ is the turbulent kinetic energy and $\eta$ is the index of the power-law decay.
For incompressible HD turbulence, $\eta$ is $-10/7$ for Loitsyansky-type \citep{Proudman1954} and $-6/5$ for Saffman type \citep{Saffman1967} spectra \citep[see also][]{Ishida2006, Davidson2010}. Compared to it, MHD turbulence still exhibits rapid decay with characteristic $\eta\sim -1$. For example, \citet{mac1998Ekdecay} reported $-1.2<\eta<-0.85$ for non-self-gravitating compressible MHD and HD turbulence. Regarding incompressible MHD, \citet{Biskamp1999} and \citet{Campanelli2004} showed that the initial magnetic helicity does not affect the decay index, which remains around $-1$.
For isothermal compressible MHD turbulence, \citet{Christensson2001} obtained a similar result of $\eta \sim-1.1$ and also observed that the initial magnetic helicity had no effect. 
Even when considering different equations of state, by $P\propto\rho^\gamma$ (with $P$ being pressure, $\rho$ density, and $\gamma$ the polytropic index varying between 0.7 and 5/3),  \citet{Lim2020} still found $\eta \sim -1$. Recent work also indicates that $\eta\sim -1$ holds for magnetic reconnection controlled \citep{Hosking2021} or shock-driven \citep{Hew&Federrath2023} MHD turbulence.
While numerous studies have characterized MHD turbulence decay in non-self-gravitating systems, investigations incorporating self-gravity remain exceptionally limited. \citet{ostriker2001density} have reported for self-gravitating MHD turbulence no significant difference compared to non-gravitating cases ($\eta\sim -1$), suggesting that self-gravity does not fundamentally alter the global decay rate. Importantly, all these studies estimated energy decay from the entire simulation domain. We wonder whether turbulence decays uniformly across all density regimes or exhibits distinct behavior in high-density regions, such as prestellar cores, which directly relate to star formation.
Given that high-density cores constitute a minor portion of the cloud in terms of both volume and mass, domain-averaged analysis is not able to reveal distinct core behaviors, if there are any.

Following \citet{ostriker2001density}, we adopt the parameter $\eta$ to quantify the influence of self-gravity on turbulence. We have taken an additional step to investigate how this effect varies with density. While following earlier work in referring to $\eta$ as the \textit{decay rate}, we thank the referee for pointing out that the $\eta$ influence by self-gravity (a driving term in the momentum equation) should more accurately be termed as \textit{rate of change} to encompass both driving and decay, rather than just a decay rate.
In Section \ref{sec:simulation_setting}, we describe the simulation suite, including gravitating and non-gravitating cases. Section \ref{sec:method} details the spatial gridding and density-binning techniques. 
We employ two complementary methods: a linear regression analysis and the traditional power-law fit to measure the change of turbulence level in Section \ref{sec:results}.
In Section \ref{sec:Discussion}, we conduct hypothesis tests to validate the robustness of the findings discussed in Section \ref{sec:results}. Additionally, we provide a comparison of our work with the existing literature in this section.
% Results in Section \ref{sec:results} demonstrate (1) the consistency of our $\eta$ with prior work and (2) density-dependent deviations in decay rates both from linear and power-law fits. These findings are subjected to hypothesis testing in Section \ref{sec:permutation}. Comparisons with previous literature and relevant observations are summarized in Sections \ref{sec:comparison decay rates} and \ref{sec:relevant_observations}, respectively. The discussion on gravitational collapse and velocity dispersion is provided in Section \ref{sec:high_density_not_collapse}.

\section{MHD simulations}
\label{sec:simulation_setting}
We adopt the simulation presented in \citet{cao2023turbulence} and \citet{zhang2019anchoring}, which accurately replicates the properties of the observed molecular clouds.
These properties include the presence of \textit{ordered cloud magnetic fields} \citep{li2009anchoring,li2015self, PlanckXXXV2016}, notably \textit{deviated core magnetic fields} \citep{Zhang2014, Hull2014,zhang2019anchoring}, a spatial magnetic field strength versus volume density relation (spatial $B$-$n$ relation) with an index of 2/3 \citep{crutcher2010, Jiang2020}, magnetic criticality values ranging from 1 to 2 for cores \citep{li2013bimodal,li2015self, MyersandBasu2021}, and core density profiles following $n\propto r^{-1.46\pm 0.12}$ \citep{Pirogov2009, KuronoApj2013, Lombardi2010, zhang2019anchoring}. 
To the best of our knowledge, no other published cloud simulations have attained as comprehensive a replication of observations.
We summarize the simulation setup here; for a more comprehensive understanding, please refer to \citet{cao2023turbulence} and \citet{zhang2019anchoring}.

The simulation domain is a periodic cube with a scale $L$ of 4.8 pc, resolved uniformly by $960^3$ voxels. The initial condition comprises a molecular cloud with a uniform initial density of $ n_{\text{ini}} = 300$ H$_2$ $\text{cm}^{-3}$ in an isothermal state of $ T = 10 \text{ K} $, with a mean molecular weight of 2.36. An initial magnetic field of 14.4\,$\mathrm{\mu G}$ uniformly permeates the whole cloud along $z$-direction.
As a result, the sound speed $ c_s $ and the ratio of thermal pressure to magnetic pressure $ \beta $ are $187.7$ m s$^{-1}$ and $ 0.05 $, respectively. The free-fall time initially is about 2 Myr.

Two MHD simulations are performed using ZEUS-MP \citep{hayes2006zeusMP}, with or without self-gravity. Each simulation is split into two distinct phases, with phase I dedicated to turbulence driving and phase II focusing on turbulence decay.
In the first phase, turbulence is driven in a divergence-free manner (purely solenoidal), injecting energy of $c_s^2/2$ per unit mass every 0.01 Myr at a scale of $L/2$, following the turbulence driving scheme used by \citet{Otto_2017}.
Upon reaching turbulence energy saturation at $3.2$ Myr, the cloud has a density-weighted RMS Mach number of $ \mathcal{M}_{\text{rms}} = 5.55$, which is typical of observed values \citep{larson1981turbulence,ossenkopf2002turbulent,heyer2004universality}, and an Alfv\'enic Mach number of $ \mathcal{M}_{\text{A}} = 0.60 $, which is in line with observational constraints \citep{li2009anchoring, li2015self,zhang2019anchoring}.

Two simulations differ in phase II. Following the turbulence saturation, we stop the turbulence driving and enable self-gravity in one simulation, while self-gravity remains disabled in the other.
This phase lasts 2 Myr with turbulence decaying and ends before any core could violate the Truelove criterion \citep{truelove1997jeans} in the self-gravitating one. Snapshots are captured at intervals of 0.02 Myr for temporal analysis.
At the end of this phase, after a simulated time $5.2$ Myr, the self-gravitating cloud has $ \mathcal{M}_{\text{rms}}= 3.80 $ and $ \mathcal{M}_{\text{A}} = 0.44 $. For non-self-gravitating one, the cloud has $ \mathcal{M}_{\text{rms}} = 3.73$ and $\mathcal{M}_{\text{A}} = 0.46 $.

\section{Spatial gridding and density binning method}
\label{sec:method}

This work aims to study density-dependent decay rates of turbulence, specifically in terms of velocity dispersion $\sigma_v$ and turbulent kinetic energy $E_{\sigma_v}$. We have developed a two-step algorithm to extract $\sigma_v$ and $E_{\sigma_v}$ evolution from the simulation datacube.

Firstly, we perform spatial gridding on the entire simulation cube.
We select a sub-cube scale $L_s$ of 0.3 pc, between the turbulence driving scale $L/2$ and the numerical dissipation scale ($L/50$; approximately 0.1 pc). The three-dimensional cloud domain is then divided into sub-cubes, with the total number of sub-cubes being $L^3/L_s^3$. These sub-cubes are evenly spaced and do not overlap spatially. Through spatial gridding, we can calculate $\sigma_v$, the number density of the molecules $n$, and $E_{\sigma_v}\equiv n\sigma_v^2$ within each sub-cube region for any snapshot in simulated time (omitting the atomic mass and one-half constants in $E_{\sigma_v}$ for simplicity).

\begin{figure}
	\centering
	\includegraphics[width=0.9\linewidth]{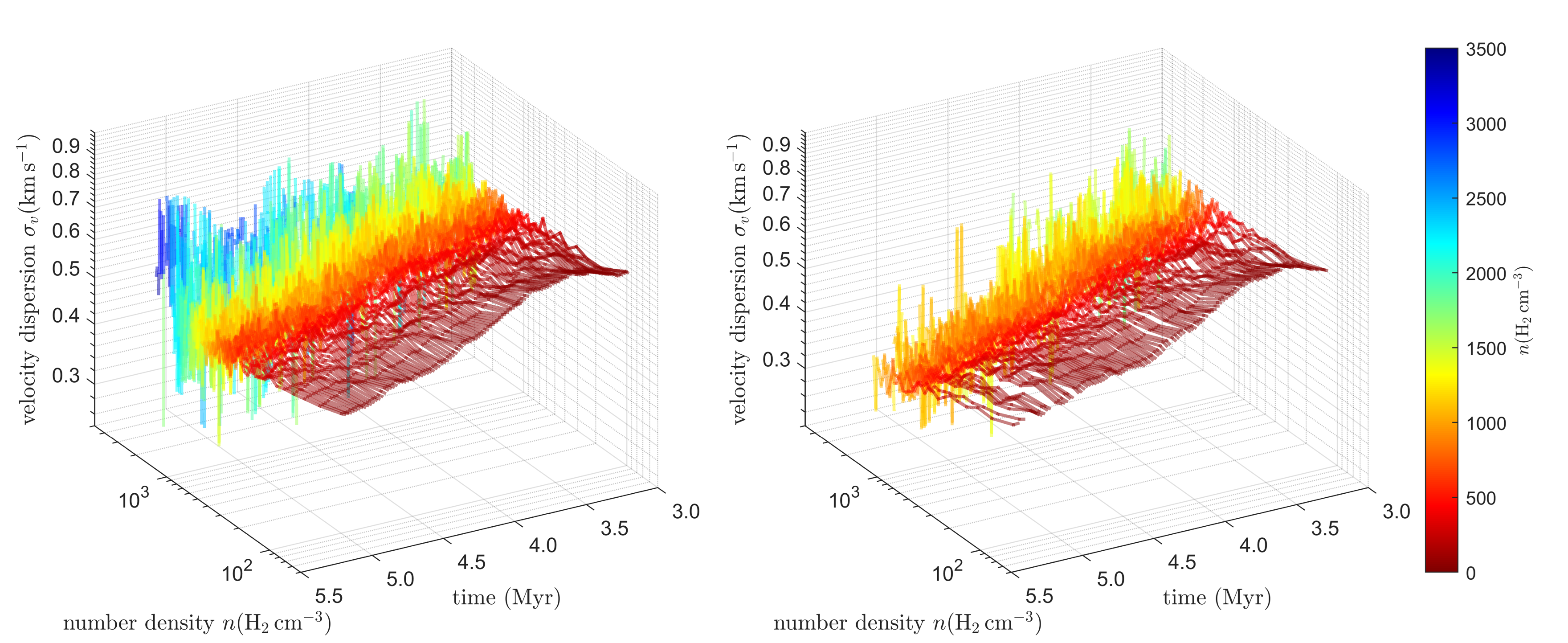}
	\caption{The waterfall plot depicts the decay of turbulence with self-gravity (left panel) and without self-gravity (right panel). Each line perpendicular to the time axis consists of the binned $\sigma_v$-$n$ pairs for a single time snapshot. In the high-density regions, $\sigma_v$ exhibits significantly different fluctuations between the two cases, whereas, in low-density regions, it decays steadily and similarly. }
	\label{fig:waterfall_sigma_n}
\end{figure}

Next, we apply density binning to these sub-cubes. 
We bin $\sigma_v$ and $E_{\sigma_v}$ of sub-cubes according to the density values $n$ within width $\Delta n$, i.e., within a specific density range $[n - \Delta n/2, n + \Delta n/2)$.
Typically, we use a bin width $\Delta n=34.9$ H$_2$ cm$^{-3}$, though this choice does not significantly impact the results. 
Post-process analyses, such as averaging and decay rate derivation, can then be applied to these density bins.
For example, the overall evolution of $\sigma_v$ with and without self-gravity is demonstrated by averaging the collected $\sigma_v$ values by \textit{waterfall plot} as shown in Figure \ref{fig:waterfall_sigma_n}.
This plot distinctly illustrates the temporal changes in the 3D phase space, highlighting the differences between self-gravitating and non-self-gravitating cases, particularly in high-density regions and later stages of the simulations.
Based on these density bins, we can derive the decay rates dependent on density, which are detailed below along with the results.

\section{Results}
\label{sec:results}

\begin{figure}
    \centering
    \includegraphics[width=0.9\linewidth]{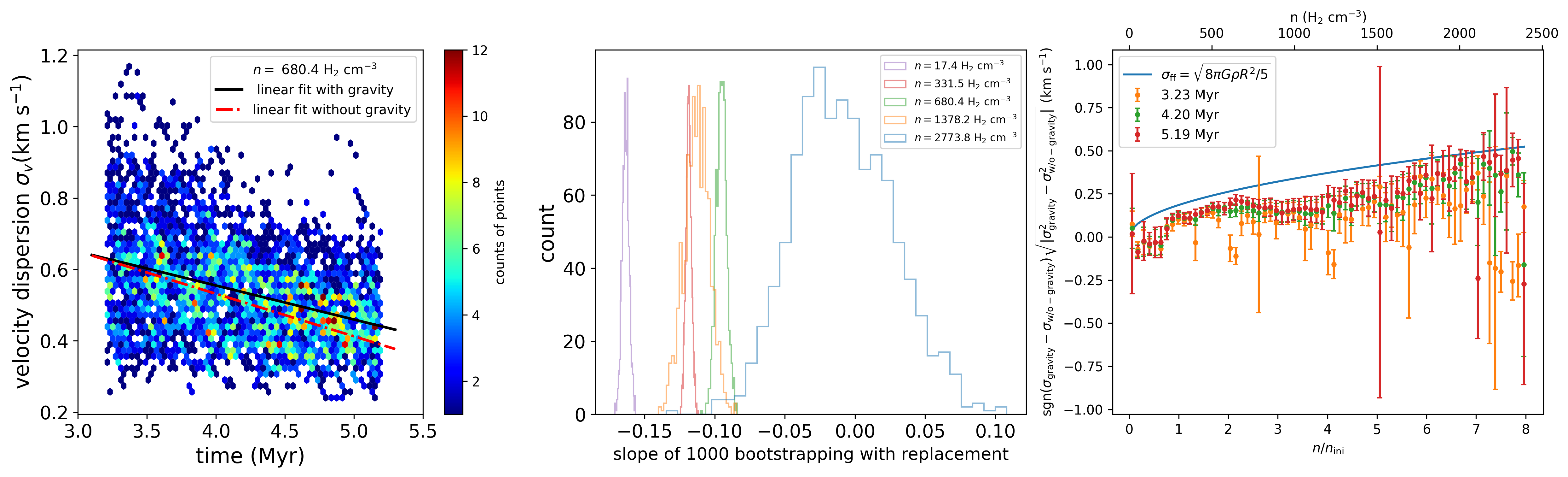}
    \caption{
    \textbf{Left}: hexagonal binning plot of self-gravitating sub-cubes in a specific density bin ($680.4$ H$_2$ cm$^{-3}$). The color indicates the point count in each hexagon. 
    The plot begins at 3.2 Myr when the turbulence saturated and driving ceased. 
    The black line shows the linear fit result determined by the median of the distribution from the bootstrap method with replacement (middle panel).
    While the corresponding hexagonal plot for the non-self-gravitating case is not displayed, its fitting result is represented by the red dash-dot line for comparison. The differences between these two fits contribute to the right panel.
    \textbf{Middle}: histograms of fitted slopes ($k_\sigma$) from five sampled density bins. Each histogram includes 1000 bootstrap replicates with replacement, divided into 30 bins. 
    The distributions give the main values (median) and uncertainties (IQR).
    \textbf{Right}: the velocity dispersion difference $\mathrm{sgn}(\sigma_\text{gravity} - \sigma_\text{w/o-gravity})\sqrt{|\sigma_\text{gravity}^2 - \sigma_\text{w/o-gravity}^2|}$, where $\mathrm{sgn}(x)$ is the sign function. $\sigma_\text{gravity}$ and $\sigma_\text{w/o-gravity}$ are the estimated $\sigma_v$ values in the self-gravitating and non-self-gravitating density bins, respectively. These estimates are derived from the linear fit $\sigma_v = k_\sigma t + b$ for a specific time $t$. The uncertainties in $k_\sigma$ and the intercept $b$ are obtained from the IQR of their respective bootstrap distributions. Then the uncertainties in the velocity dispersion difference are obtained by the error propagation. 
    The blue line shows the free-fall velocity dispersion ($\sigma_\text{ff}$) for comparison. 
    }
    \label{fig:bootstrap_replacement}
\end{figure}

The decay rates of $\sigma_v$ and $E_{\sigma_v}$ within each density bin are derived through the bootstrap method \citep{hesterberg2011bootstrap} and linear fit. The left panel of Figure \ref{fig:bootstrap_replacement} is an illustration of $\sigma_v$ decay in the $n = 680.40\pm17.45$ H$_2$ cm$^{-3}$ density bin, where the color map illustrates the clustering density of scatter points. 
Taking this as an example, we briefly describe the derivation process based on \citet{hesterberg2011bootstrap}. We randomly draw scatter points with replacements to form the bootstrap replicate. The replicate should contain the same number of points as the original corresponding density bin, permitting repeated draws of the same data points.
We then perform linear fits on these replicates to obtain the slopes. Once 1000 slopes are collected, we can obtain a distribution in the middle panel of Figure \ref{fig:bootstrap_replacement}, as well as distributions derived from other density bins for comparison. 

\begin{figure}
    \centering
    \includegraphics[width=0.9\linewidth]{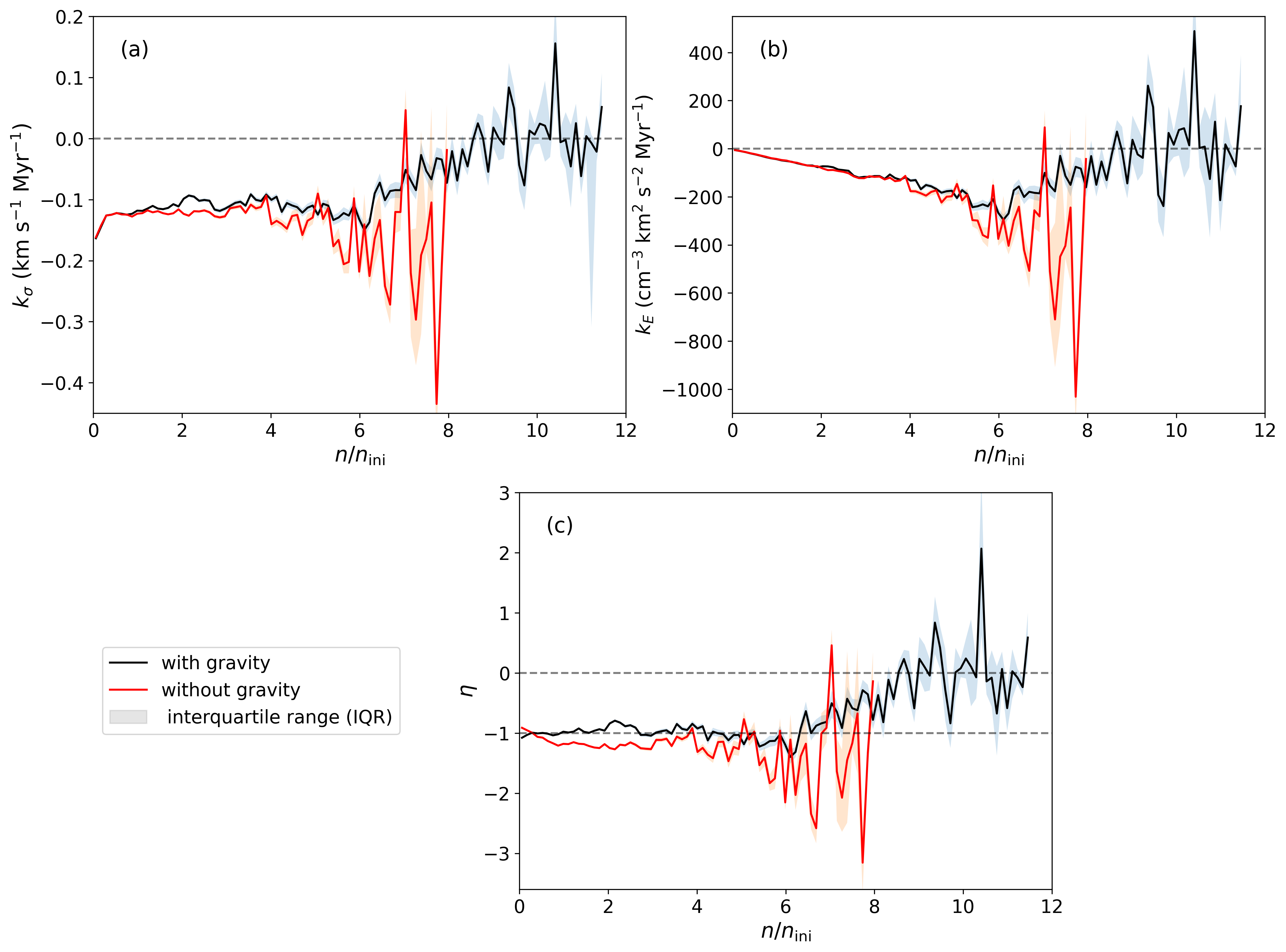}
    \caption{A comparison plot of the self-gravitating (black) and non-self-gravitating (red) turbulence decay rates versus density for $\sigma_v$ (a), $E_{\sigma_v}$ (b) and power index $\eta$ (c), respectively. 
     The median values (solid lines) are used to avoid the bias of the skewed distribution. The shaded areas illustrate the corresponding interquartile ranges. The dash lines are for visual guidance. }
   \label{fig:decay_rates_comparison}
\end{figure}

We define the $\sigma_v$ decay rate $k_\sigma$ using the median of the distribution, rather than the mean, due to the skewness, especially in higher density bins. The uncertainty is determined by the interquartile range (IQR) of the distribution.
The derivation of $E_{\sigma_v}$ decay rate $k_E$ has the same process except replacing the $\sigma_v$ scatters with $E_{\sigma_v}$ scatters from density binning.
By extending such a process to all the density bins, 
one can obtain the correlation between decay rate and density, for $\sigma_v$ (Figure \ref{fig:decay_rates_comparison}a) and $E_{\sigma_v}$ (Figure \ref{fig:decay_rates_comparison}b), respectively.

For the case with self-gravity (black line in Figure \ref{fig:decay_rates_comparison}), density regimes below $6n_{\text{ini}}$ exhibit $k_\sigma$ around $-0.1$\kmspMyr, whereas higher density regimes display $k_\sigma\sim 0$. This pattern underscores that, in self-gravitating MHD turbulence, turbulence at high-density regimes experiences a slower decay than in low-density regimes. Consequently, this slower decay contributes to the elevated levels of velocity dispersion observed in the left panel of Figure \ref{fig:waterfall_sigma_n}, i.e., the blue part in the figure. On the contrary, the non-self-gravitating case (red line) in Figure \ref{fig:decay_rates_comparison} (a) does not have a near-zero $k_\sigma$ in high-density regions.

In light of the significant density dependence shown for $k_\sigma$, we shall also anticipate density dependence for $k_E$. Due to the density component's contribution, $E_{\sigma_v}$ in low-density regimes would decay slower than the intermediate-density regimes, as illustrated by the negative slope in both self-gravitating (black line) and non-self-gravitating (red line) $k_E$ to the left of $6n_{\text{ini}}$ in Figure \ref{fig:decay_rates_comparison} (b). As for the right of $6n_{\text{ini}}$ in Figure \ref{fig:decay_rates_comparison} (b), with the self-gravitating $k_\sigma$ approaching zero, the associated $k_E$ also diminishes to near-zero levels. However, the decay rate of $E_{\sigma_v}$ in the non-self-gravitating case does not approach the zero level in the high-density regimes. 
Consequently, $k_E$ of the self-gravitating cloud exhibits a V-shaped pattern as a function of gas density, as the black line shown in Figure \ref{fig:decay_rates_comparison} (b), while the non-self-gravitating one (red line) lacks.

In addition, previous studies have commonly utilized a power-law model to describe the decay of $E_{\sigma_v}$ within the whole simulation domains \citep[e.g.,][]{mac1998Ekdecay, Hew&Federrath2023}. 
The power law fit takes the form \(E_{\sigma_v} = E_0 (1+t/t_0)^\eta\), where \(E_0\) represents the initial energy, \( t_0 \) is the sound crossing time ($t_0 = 1.56\,\text{Myr}$ in this work.), and \( t \) is the time elapsed since the initiation of turbulence decay. 
The validity of this power law stems from the large-scale spectrum's dominance in total energy, as discussed by \citet{Galtier1997}.
However, this assumption is less applicable to small sub-cubes, where the decay appears to be more linear, as indicated by the left panel of Figure \ref{fig:bootstrap_replacement}. 
Nevertheless, for historical and comparison purposes, we also perform a power-law fit (Figure \ref{fig:decay_rates_comparison}c) to compare with established indices $\eta$ in the literature.
Besides agreeing with Figure \ref{fig:decay_rates_comparison} (a) on the bimodal decay rate as a function of density, the index ($\eta$) of the power-law fit on low-density regions also aligns with the values reported by \citet{mac1998Ekdecay}, who derived a single decay rate (power law) from their simulation without discriminating based on density.
This is reasonable, as the majority of a cloud's mass exists in low-density regions. Please refer to the discussion in section \ref{sec:comparison decay rates}.

\section{Discussion}
\label{sec:Discussion}

\subsection{Permutation Test}
\label{sec:permutation}

The decay rates in Figure \ref{fig:decay_rates_comparison} exhibit fluctuations, especially in the regions $n>6n_{\text{ini}}$, which necessitates an examination of the robustness of the findings discussed in the preceding section. Due to the unknown distribution of the small population, we utilized a non-parametric permutation test \citep{phipson2010permutation} using the SciPy package \citep{2020SciPy-NMeth}. 

Firstly, we generate independent samples by bootstrapping. We randomly select one value of $k_\sigma$ from each density bin, exemplified by the histograms in the middle panel of Figure \ref{fig:bootstrap_replacement}. 
Utilizing the $6n_{\text{ini}}$ as the dividing point, we divide the density bins into two groups, noted as the lower-density ($n<6n_{\text{ini}}$) group and higher-density ($n>6n_{\text{ini}}$) group. 
The averaged $k_\sigma$ of the two groups are denoted by $\mu_\text{low}$ and $\mu_\text{high}$, respectively.
For reference, if we use the median values instead of the random selection (as indicated by the solid lines in Figure \ref{fig:decay_rates_comparison}), the self-gravitating case exhibits $\mu_\text{low} = - 0.11$\kmspMyr and $\mu_\text{high} = -0.03$\kmspMyr. In contrast, the non-self-gravitating $k_\sigma$ shows $-0.13$\kmspMyr and $-0.18$\kmspMyr.

Then the $k_\sigma$ values are randomly selected to form group 1, which shares the same size as the lower-density group; the remaining $k_\sigma$ values form group 2. Their averages are $\mu_1$ and $\mu_2$, respectively. The $k_\sigma$ sampling process above is repeated 10,000 times.
We use the difference between the two group means, $\mu_1 - \mu_2$, as the test statistic of the permutation test.

Our null hypothesis $H_0$ states that the two groups have the same distribution, implying that $\mu_\text{low} = \mu_\text{high}$. The alternative hypothesis $H_1$ suggests that $\mu_\text{low} < \mu_\text{high}$. It is important to note that $\mu_\text{low} < 0$ and $\mu_\text{high}\lesssim 0$.
The probability of obtaining $\mu_1 - \mu_2 \leq \mu_\text{low}-\mu_\text{high}$ under the null hypothesis is defined as the p-value. In other words, the p-value quantifies the likelihood that the observed decay rate difference in Figure \ref{fig:decay_rates_comparison} (a) could arise from a random event, assuming the null hypothesis is true.
We recommend that readers refer to the SciPy implementation \footnote{\href{https://docs.scipy.org/doc/scipy/reference/generated/scipy.stats.permutation_test.html}{https://docs.scipy.org/doc/scipy/reference/generated/scipy.stats.permutation\_test.html}} and our codes \footnote{CUHK Data repository \href{https://doi.org/10.48668/KDQO4F}{https://doi.org/10.48668/KDQO4F}} for further details.

\begin{figure}
    \centering
    \includegraphics[width=0.9\linewidth]{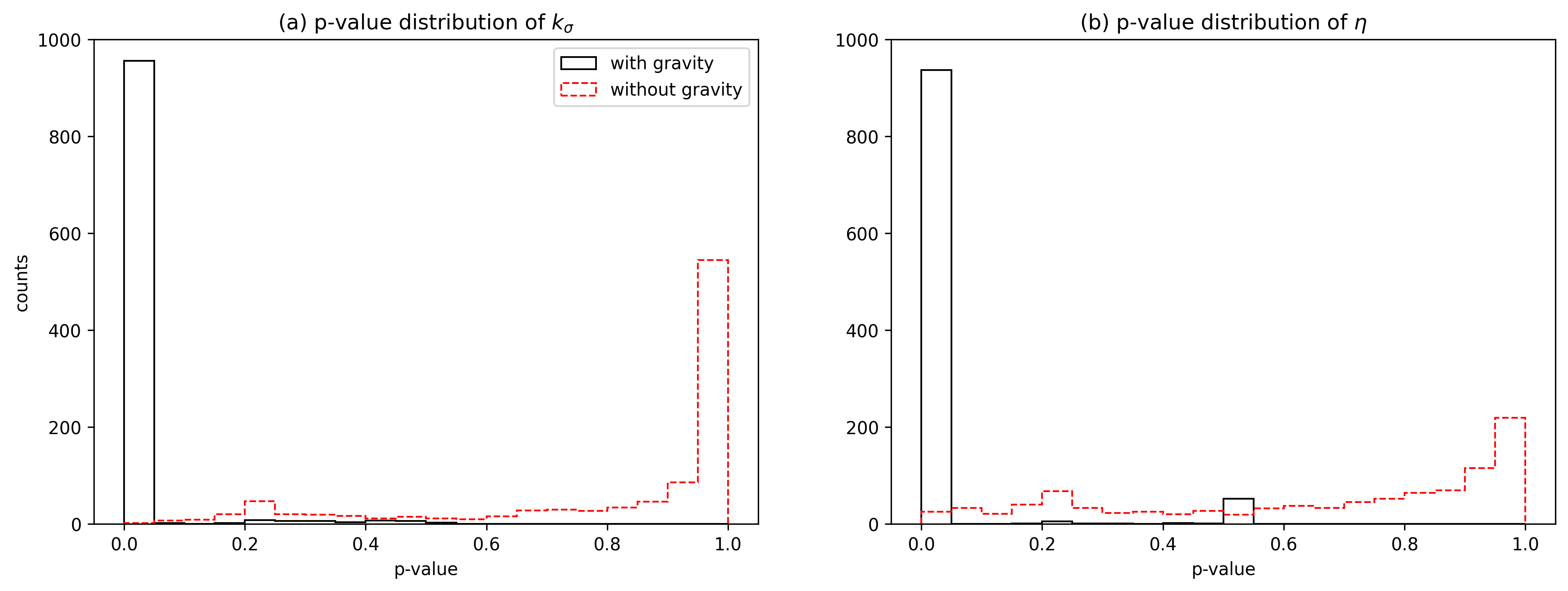}
    \caption{Histograms of the statistical significance (p-value) from permutation tests of $k_\sigma$ (a) and $\eta$ (b), for self-gravitating and non-self-gravitating simulation data, respectively.}
    \label{fig:permutation_results}
\end{figure}

Traditionally, a p-value less than 0.05 is considered significant evidence against the null hypothesis and in favor of the alternative. 
We repeated the above analysis 1000 times. $H_0$ was rejected 956 times for the self-gravitating case and only 2 times for the non-self-gravitating case. 
Similarly, when we replace $k_\sigma$ by $\eta$, $H_0$ was rejected 937 and 25 times, respectively, with and without gravity. Figure \ref{fig:permutation_results} illustrates these permutation results, which align with our earlier visual observations in Section \ref{sec:results}.

\subsection{The comparison with earlier works}
\label{sec:comparison decay rates}

For the low-density regimes ($n<6 n_{\text{ini}}$) in Figure \ref{fig:decay_rates_comparison} (c), we find $\eta \simeq -1.0$ in the gravitating case and $\eta \simeq -1.2$ in non-gravitating case, consistent with $-1.2<\eta<-0.85$ for non-self-gravitating isothermal magneto-turbulence in \citet{mac1998Ekdecay} and self-gravitating models with $\eta \simeq-1$ in \citet{ostriker2001density} and \citet{Tilley2007}. It can be explained by the prevalence of low-density regions, which make up a large fraction of both the volume and mass of the molecular cloud.  For a specific example, in the simulation with self-gravity, regions with densities less than or equal to $6n_\text{ini}$ make up to 97\% of the cloud's volume and 70\% of its mass during the decaying phase. Thus, the molecular cloud can be considered low-density dominated. It may be worth noting that \citet{mac1998Ekdecay} used the same algorithm (ZEUS code) in their models, but with lower resolution, despite having examined spatial (grid) convergence.

In the high-density regimes of our self-gravitating simulations, we observe $\eta \simeq 0$, affirming that regimes with $n\ge 8n_{\text{ini}}$ exhibit turbulent velocity and energy decay rates close to 0 in Figures \ref{fig:decay_rates_comparison}. The prevalence of low-density regions buried such turbulence characteristics of high-density regions in earlier works.

\subsection{Relevant observations}
\label{sec:relevant_observations}

While the direct observation of turbulence dissipation rates across various densities remains challenging, an increasing trend of the Alfv\'en Mach number $ \mathcal{M}_{\text{A}} $ with density has been observed (e.g., Figure 2f of \citealt{pattle2022magnetic}).

Density not only directly contributes to the increase in $\rho \sigma_v^2$ through the $\rho$ term but also influences $\sigma_v$ itself, due to the density-dependent $k_\sigma$ observed in this work. Particularly in the late stage of phase II, the averaged $\sigma_v$ exhibits a proportionality to density, attributable to the near-zero $k_\sigma$.
Overall, $E_{\sigma_v}$ shows a positive correlation with density. The rate of increase in turbulent kinetic energy with density is observed to be five to ten times higher than that of the magnetic field energy across the time snapshots.
The corresponding increase rates for the magnetic field energy, as inferred by B-field strength in Figures 2 and 3 of \citet{cao2023turbulence}, will be further elaborated in our forthcoming paper. Such disparity in these energy increase rates yields the rise of $\mathcal{M}_{\text{A}}$ with density, leading to the formation of super-Alfv\'enic high-density regions. 

The super-Alfv\'enic cores seem to contradict another observation: quiescent cores. However, it is important to note that turbulence velocities are scale-dependent. Quiescent cores are usually observed as high-density regions with sizes well below 0.1 pc, as reported in \citet{FullerMyers1992ApJ, Goodman1998ApJ, Choudhury2021, WangWang2023, Barnes2023motherofdragons}. Here, we demonstrate that at scales below 0.1 pc, turbulence can also be sub-sonic in our simulations.

Due to the proximity of 0.1 pc (20 pixels) to the numerical dissipation scale of the current simulation, we can only extrapolate the turbulence power law from scales above 60 pixels to estimate velocity dispersion at the 20-pixel scale.

\begin{figure}
	\centering
	\includegraphics[width=0.5\linewidth]{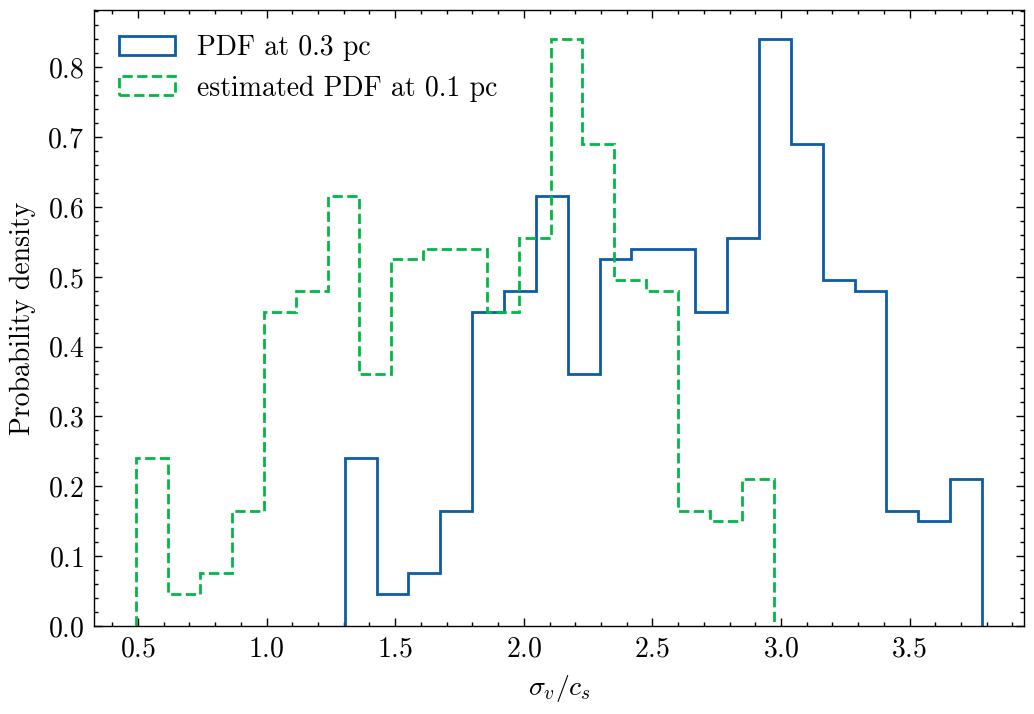}
	\caption{The probability density functions of turbulence Mach numbers $\mathcal{M}$ are depicted for sub-cubes surpassing the density threshold $n>6n_{\text{ini}}$. The solid line represents the histogram of all the eligible sub-cubes spanning from 4.59 Myr to 4.99 Myr, with a total of 21 snapshots in phase II, facilitating a comprehensive distribution display. The dashed line is the distribution extrapolated by the $\alpha=-5/3$ power law from the solid line.}
	\label{fig:quiscentMachPDF}
\end{figure}

Figure \ref{fig:quiscentMachPDF} shows the distribution of $\mathcal{M}$ from all sub-cubes ($60^3$ pixels or $0.3^3$ pc$^3$) with densities above $6n_\text{ini}$, spanning a period from 4.59 to 4.99 Myr. All sub-cubes exhibit super-sonic turbulence. To estimate the Mach numbers of $0.1^3$ pc$^3$ subregions within each sub-cube, we assume an energy spectrum index of -5/3 (corresponding to a velocity spectrum index of -1/3), as proposed by \citet{K41}. The distribution of these estimated Mach numbers is also shown in Figure \ref{fig:quiscentMachPDF}. Notably, the lower tail of this distribution can be sub-sonic. It is important to note that if a Burgers spectrum is assumed (with an energy index of -2; \citealt{Burgers1974, heyer2004universality, zhang2019anchoring}), the Mach numbers for 0.1-pc cores could be even more sub-sonic.

\subsection{Enhanced turbulence without significant collapsing}
\label{sec:high_density_not_collapse}
It is common to encounter simulations or observations that associate the contraction of clouds with internal kinetic energy (e.g., \citealt{BallesterosParedes2011a, BallesterosParedes2011b,Seifried2017, BallesterosParedes2018,IbanezMejia2016, IbanezMejia2022}). However, it is important to note that the dense regions in our simulations, which display enhanced turbulence (as illustrated in Figure \ref{fig:decay_rates_comparison}, exceeding six times $n_\text{ini}$), are mostly not collapsing.  \citet{cao2023turbulence} showed that only one of the three densest regions is undergoing contraction, while the other two are either stable or expanding. They offer a comprehensive analysis of how magnetic, thermal, gravitational, and turbulent forces interact to explain these phenomena. 

From another angle, $\sigma_v$ in our simulation significantly exceeds the standard free-fall velocity dispersion $\sigma_\text{ff}\equiv\sqrt{8\pi G\rho R^2/5}$ \citep{BallesterosParedes2018}, where \(R\) is approximated by the sub-cube length \(L_s\), \(\rho\) is the mean density of the sub-cube, and \(G = 4.4998 \times 10^{-3}\) pc\(^3\) Myr\(^{-2}\) M\(_{\odot}\) is the gravitational constant. $\sigma_\text{ff}$ is the maximum velocity dispersion derived from the potential energy. 
This further suggests that the velocity dispersion in our simulation is not due to core collapse. It is also essential to understand that gravitational potential energy is determined solely by mass distribution; therefore, any increase in density, such as that caused by turbulence, can transform potential energy into kinetic energy. Collapse is not the only mechanism that can contribute to $\sigma_\text{ff}$, regardless of its designation. 

Additionally, Figure \ref{fig:bootstrap_replacement} (right panel)  demonstrates that $\sigma_\text{ff}^2$ indeed establishes the upper limit of the turbulent energy difference between scenarios with and without self-gravity. For instance, from the left panel of Figure \ref{fig:bootstrap_replacement}, we can derive a $\sigma_v^2$ decay function for a specific density. We can calculate the difference between $\sigma_v^2$ with and without gravity at a given time. The right panel of Figure \ref{fig:bootstrap_replacement} presents examples of the signed square roots of these $\sigma_v^2$ differences, with the upper limit defined by $\sigma_\text{ff}$. This clearly illustrates the increase of turbulence energy in high-density regions resulting from the potential energy released due to gas concentration.

\section{Conclusions}

Our research, leveraging spatial gridding and density binning techniques on MHD simulations, is the first to study the density-dependent turbulence decay rate.
Our key finding is that only self-gravitating MHD turbulence exhibits a density-dependent decay pattern for velocity dispersion and turbulent energy. 
While maintaining the characteristic domain-averaged $\eta\sim-1$, dense regions exhibit significantly reduced $\eta\simeq 0$.
Specifically, in regions with densities above 1800 H$_2$ cm$^{-3}$, the rate of change of turbulence velocity is significantly slower, with a mean value of $-0.03$\kmspMyr, compared to $-0.11$\kmspMyr in regions with densities below 1800 H$_2$ cm$^{-3}$.
This contrast persists through rigorous statistical validation via permutation tests with bootstrap error propagation.
This slower decay indicates that turbulence in high-density regions persists longer than previously thought and assists in addressing the decades-long question regarding the persistence of cloud turbulence.

\section*{Acknowledgments}

We thank Dr. Zhuo Cao for the insightful and comprehensive discussions about numerical simulation. Special thanks are also due to Professor Xiaodan Fan for the expert guidance provided on statistical aspects of the research. Shibo Yuan acknowledges Xinrui Yu for her valuable contributions to the discussions about permutation tests.
This research is supported by General Research Fund grants from the Research Grant Council of Hong Kong (14306323, 14307222, and 14303921) and by a Collaborative Research Fund grant (C4012-20E). 

%%%%%%%%%%%%%%%%%%%%%%%%%%%%%%%%%%%%%%%%%%%%%%%%%%
\section*{Data Availability}

The data used for supporting the findings of this study are available from CUHK Research Data Repository, \href{https://doi.org/10.48668/KDQO4F}{https://doi.org/10.48668/KDQO4F}.



\bibliographystyle{mnras}
\bibliography{ref} 

%%%%%%%%%%%%%%%%%%%%%%%%%%%%%%%%%%%%%%%%%%%%%%%%%%

%%%%%%%%%%%%%%%%%%%%%%%%%%%%%%%%%%%%%%%%%%%%%%%%%%
\end{CJK*}
\end{document}